\def\year{2022}\relax
\documentclass[letterpaper]{article} 
\usepackage{aaai22}  
\usepackage{times}  
\usepackage{helvet}  
\usepackage{courier}  
\usepackage[hyphens]{url}  
\usepackage{graphicx} 
\usepackage{natbib}  
\usepackage{caption} 
\DeclareCaptionStyle{ruled}{labelfont=normalfont,labelsep=colon,strut=off} 
\usepackage{color}  

\usepackage{paralist}
\usepackage{algorithm}
\usepackage{algorithmic}
\usepackage{amsmath} 
\usepackage{newfloat}
\usepackage{listings}
\floatstyle{ruled}
\newfloat{listing}{tb}{lst}{}
\floatname{listing}{Listing}
 \usepackage{amssymb}
 \usepackage{adjustbox}
 \usepackage{adjustbox}

\title{Deep Recurrent Neural Network with Multi-scale Bi-directional Propagation for Video Deblurring\\}

\author{
    Chao Zhu\textsuperscript{\rm 1}~\equalcontrib,
    Hang Dong\textsuperscript{\rm 2}~\equalcontrib~\thanks{Corresponding author.},
    Jinshan Pan\textsuperscript{\rm 3},
    Boyang Liang\textsuperscript{\rm 1},
    Yuhao Huang\textsuperscript{\rm 1},
    Lean Fu\textsuperscript{\rm 2},
    Fei Wang\textsuperscript{\rm 1}
}
\affiliations{
    
    \textsuperscript{\rm 1}Xian Jiaotong University\\
    \textsuperscript{\rm 2}ByteDance Intelligent Creation Lab\\
    \textsuperscript{\rm 3}Nanjing University of Science and Technology\\

}

\begin{document}

\maketitle

\begin{abstract}

The success of the state-of-the-art video deblurring methods stems mainly from implicit or explicit estimation of alignment among the adjacent frames for latent video restoration. 
However, due to the influence of the blur effect, estimating the alignment information from the blurry adjacent frames is not a trivial task.
Inaccurate estimations will interfere the following frame restoration. 
Instead of estimating alignment information, we propose a simple and effective deep Recurrent Neural Network with Multi-scale Bi-directional Propagation (RNN-MBP) to effectively propagate and gather the information from unaligned neighboring frames for better video deblurring. 
Specifically, we build a Multi-scale Bi-directional Propagation~(MBP) module with two U-Net RNN cells which can directly exploit the inter-frame information from unaligned neighboring hidden states by integrating them in different scales.
%
%
Moreover, to better evaluate the proposed algorithm and existing state-of-the-art methods on real-world blurry scenes, we also create a Real-World Blurry Video Dataset (RBVD) by a well-designed Digital Video Acquisition System (DVAS) and use it as the training and evaluation dataset. 
Extensive experimental results demonstrate that the proposed RBVD dataset effectively improves the performance of existing algorithms on real-world blurry videos, and the proposed algorithm performs favorably against the state-of-the-art methods on three typical benchmarks.
The code is available at \emph{https://github.com/XJTU-CVLAB-LOWLEVEL/RNN-MBP}.

\end{abstract}

\section{Introduction}
%
With the rapid development of smart-phone, live-streaming businesses and cloud computing services, videos become one of the most popular communication mediums in our daily life, 
plenty of people enjoy using hand-held devices to record their moments and share online. 
However, the captured videos often suffer from motion blur effect which is usually caused by camera shaking and object moving.
Therefore, how to remove the motion blur in the videos becomes an urgent issue in both the academic and industrial communities.

The goal of video deblurring algorithm is to recover a sharp video from a blurry one, which is highly ill-posed.
Lots of traditional methods~\cite{4409009, 6909750, 7299181, 2014Modeling} usually impose kinds of hand-crafted priors on both blur kernels and latent frames and alternatively estimate the blur kernels and latent frames.
However, these methods highly rely on the accuracy of kernel estimation which leads to unstable deblurring performance.
Instead of using hand-crafted priors and alternative solutions, the deep convolutional neural network (CNN), 
provides an effective way to directly estimate the clear videos from blurry ones~\cite{2018Adversarial,zhou2019davanet, Gast_2019_corr, 2018Spatio,nah2019recurrent, 2019EDVR, Pan_2020_CVPR, Kim_2017_ICCV, ESTRNN, suin2021gated}. 
We note that when restoring the clear frames, most traditional methods~\cite{Motion2008, 6909750, 7299181} and the CNN-based ones~\cite{2018Spatio, Kim_2017_ICCV, 2019EDVR, Pan_2020_CVPR} usually involve the alignment estimations by optical flow, deformable convolution, or alignment filters and so on. 
%
%
For example, EDVR~\cite{2019EDVR} and CDVD-TSP~\cite{Pan_2020_CVPR} use multi-scale deformable convolution and optical flow estimation to align inter-frame information respectively.
Intuitively, these video deblurring algorithms should outperform the image deblurring algorithms by taking advantages of both intra-frame (spatial) and inter-frame (temporal) information from the adjacent frames.
However, it is not trivial to estimate the alignment from blurry videos and most existing methods may fail due to the presence of motion blur and occlusion~\cite{xue2019video}. The inaccurate estimations of the alignment will affect the deblurred results. 
Therefore, on some popular public video deblurring dataset (e.g. GOPRO~\cite{deepdeblur}),
most existing video deblurring methods do not perform well while top three methods ({HINet~\cite{Chen_2021_CVPR}}, {MIMO-UNet~\cite{MIMOUNet}}, and {MPRNet~\cite{Zamir2021MPRNet}}) on GOPRO dataset are all image deblurring algorithms.
%

%
%
\begin{figure*}[!t] 
  \centering
  \includegraphics[width=0.9\linewidth]{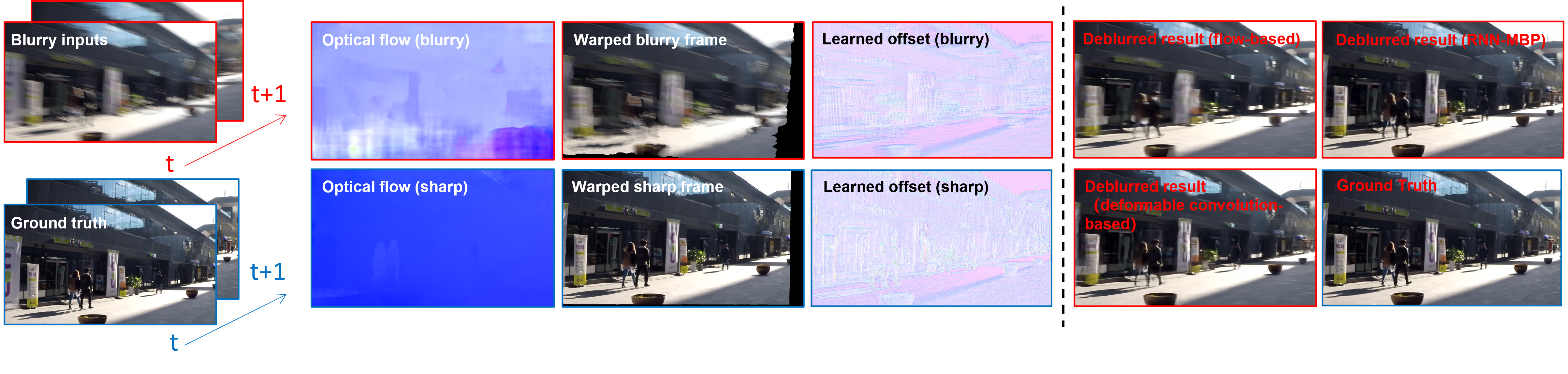}
  \vspace{-5mm}
  \caption{\textbf{
 Optical flow-based and deformable convolution-based methods on video deblurring.
 }
  We evaluate the flow-based algorithm CDVD-TSP~\cite{Pan_2020_CVPR} and deformable convolution-based algorithm EDVR~\cite{2019EDVR} using a video with significant blur from the GOPRO dataset~\cite{deepdeblur}.
  The top row shows the blurry frames, the optical flows and warped blurry frame from the CDVD-TSP, the learned offsets from the EDVR, the deblurred result of the CDVD-TSP, and the proposed deblurred result, respectively.
  The bottom row shows the sharp frames (ground truths), optical flow and warped sharp frame which are estimated from sharp frames by CDVD-TSP, the learned offsets from sharp frames by EDVR, the deblurred result of the EDVR, and ground truth result, respectively.
  Due to the influence of blur effect, the methods based on the optical flow and deformable convolution do not restore sharp frames.
  Best viewed on a high-resolution display.
  }
\label{fig:flowDeformable}
\vspace{-5mm}
\end{figure*}
%
%
%
%
Figure \ref{fig:flowDeformable} shows the intermediate and deblurred results of two video deblurring methods with explicit alignment using a video with significant blur from the GOPRO dataset. 
We feed the blurry frames and corresponding sharp frames to the flow-based method CDVD-TSP~\cite{Pan_2020_CVPR} and deformable-based method EDVR~\cite{2019EDVR}, respectively.
Note that the methods based on optical flow or the deformable convolution do not estimate the alignment information well (even though these alignment modules have been finetuned) due to the influence of the motion blur, thus leading to results with significant blur residual.
%

To overcome this problem, we propose a deep Recurrent Neural Network with Multi-scale Bi-directional Propagation (RNN-MBP) to exploit inter-frame information from neighbouring frames for video deblurring.
The proposed Multi-scale Bi-directional Propagation (MBP) module is built by two U-Net RNN cells and the bidirectional propagation scheme under a multi-scale framework. Thus, the MBP can directly gather the unaligned neighboring hidden states and exploit the inter-frame information in different scales.
%
%
%
Moreover, instead of directly reconstructing the target frame from the output of the propagation module~\cite{chan2021basicvsr}, we propose a Target Frame Re-constructor module to first fuse the features from the target frame and the multi-scale outputs from the MBP module.
By re-introducing the features from the target frame with an additional sub-network, the proposed Target Frame Re-constructor module can reinforce the importance of intra-frame information, as the intra-frame information and its importance may gradually vanish during the propagation process.
Benefiting from these designs, the inter-frame and intra-frame information features can be exploited thus facilitating the clear video restoration.
%
%

In addition, we note that most existing video deblurring algorithms are trained and evaluated on the synthetic datasets, such as GOPRO~\cite{deepdeblur} and REDS~\cite{REDS}.
Due to the domain gap between synthetic datasets and real-world ones~\cite{8953368}, the methods trained on the synthetic datasets do not generalize well on the real-world scenes as demonstrated by~\cite{lai_blur, kohler, RealBlur, ESTRNN}.
To overcome this problem, we create a Real-World Blurry Video Dataset (RBVD) by a well-designed Digital Video Acquisition System (DVAS) to evaluate the proposed method and the state-of-the-art ones.
Quantitative and qualitative experiments demonstrate that the proposed RBVD improves the performance of existing methods on real-world blurry videos, and the proposed algorithm shows better performance than the state-of-the-art methods on the RBVD test-set. 
%

The main contributions are summarized as follows:
\begin{compactitem}
    \item We propose a RNN-MBP to effectively exploit intra- and inter-frame information by directly gathering the unaligned neighboring hidden states.
    \item We propose a Target Frame Re-constructor to reinforce the importance of intra-frame information by re-introducing the features from target frames.
    \item To better evaluate different video deblurring methods on real-world blurry scenes, we create a real-world blurry video dataset by a well-designed DVAS.
    \item Both quantitative and qualitative evaluations demonstrate that our model performs favorably against the state-of-the-art methods in terms of accuracy and model size.
\end{compactitem}
%

\begin{figure*}[!t] 
    \centering
    \includegraphics[width=0.88\linewidth]{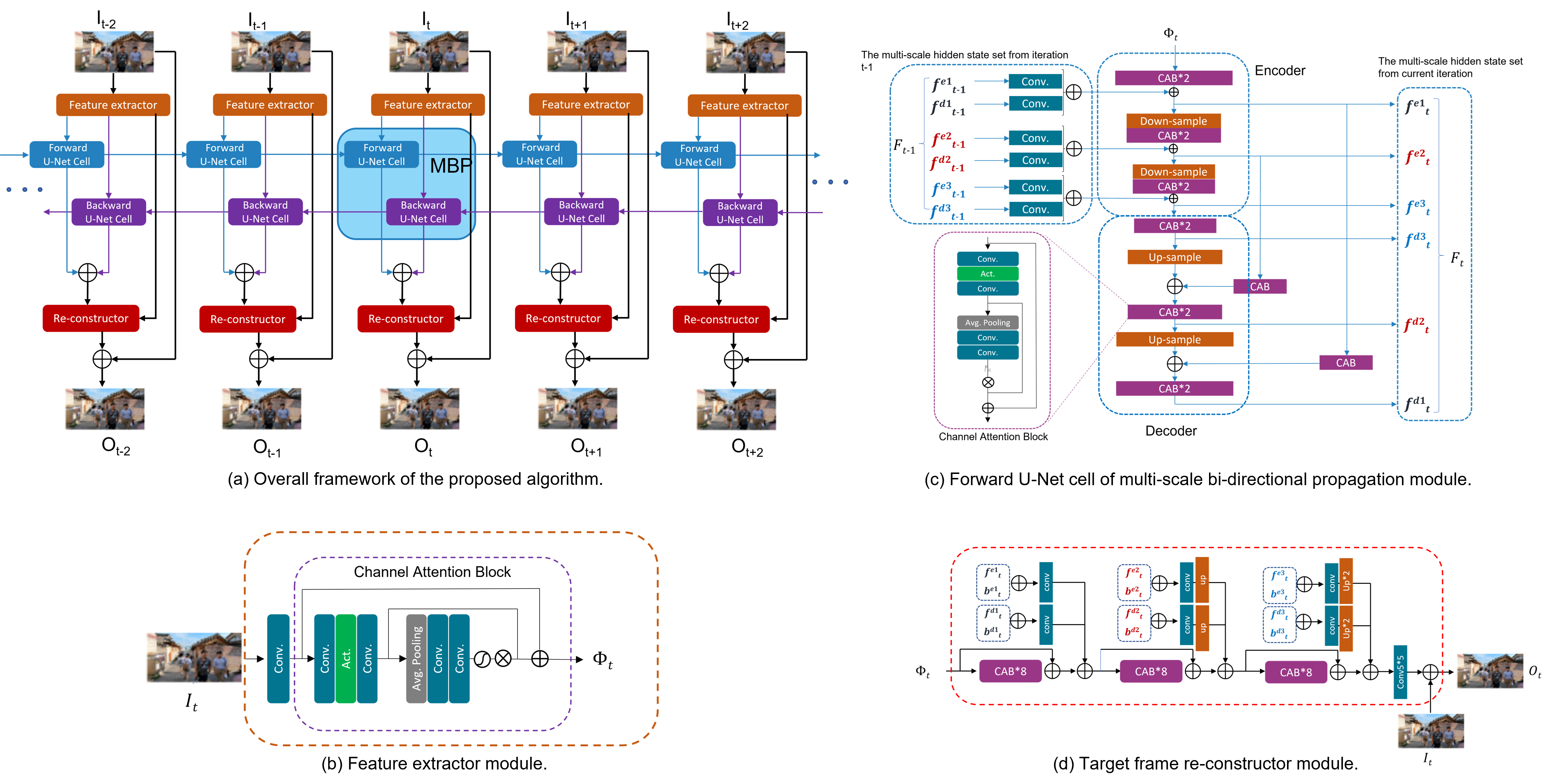}
    \vspace{-2mm}
    \caption{\textbf{Framework of the proposed RNN-MBP in video deblurring}. (a) Overall framework of the proposed algorithm. (b) Details of the Feature extractor module. (c) Details of the Multi-scale Bi-directional Propagation module. (c) Details of the Target Frame Re-constructor module. 
    The forward and backward hidden states with the same color have the same spatial scale. 
    Best viewed on a high-resolution display.}
    \label{fig:RNN}
        \vspace{-4mm}
\end{figure*}

\section{Related Work}
%
%
Significant progress has been made in single image deblurring due to the use of deep learning~\cite{Gong_2017_CVPR,Nimisha_2017_ICCV,DeblurGAN,Kupyn_2019_ICCV,Zamir2021MPRNet,Chen_2021_CVPR}. We refer readers to~\cite{2021NTIRE, 2020NTIRE, 2019NTIRE} for a comprehensive review of recent advances. In the following, we mainly briefly introduce the most related methods. 
%
%
%
%
\newline
{\bf Video Deblurring Algorithm.}
As the convolutional neural networks have achieved favorable results in image deblurring~\cite{Gong_2017_CVPR, Zhang2018, Kupyn_2019_ICCV, Zamir2021MPRNet, Chen_2021_CVPR}, researchers begin to focus on solving video deblurring tasks with deep learning techniques.
To better exploit the inter-frame (temporal) information in the adjacent frames, many implicit or explicit alignment methods are proposed.
\cite{2018Adversarial} proposes a video deblurring network which use modified 3D convolutional layers to learn spatial temporal information.
%
%
In the NTIRE19 video Deblurring Challenge~\cite{2019NTIRE}, \cite{2019EDVR} proposes a pyramid cascading deformable~(PCD) alignment module to compensate and fuse the temporal information in the feature-level which achieves the best results.
In \cite{Pan_2020_CVPR}, Pan et al. use a pre-trained PWC-Net~\cite{Sun2018PWC-Net} to estimate the optical flows to exploit the inter-frame information from adjacent frames in a cascading way.

In the meantime, 
some researchers choose the recurrent neural network~(RNN) as the framework to solve the video deblurring tasks,
as RNN is suitable for processing on the time sequential inputs.
%
\cite{Kim_2017_ICCV} develops a dynamic temporal blending layer in their spatial-temporal recurrent network which can enforce the temporal consistency among consecutive outputs.
\cite{ESTRNN} builds their RNN cells with residual dense blocks~(RDBs) and applies a global spatio-temporal attention module to fuse features from adjacent frames.
However, due to the difficulty in estimating the alignment from blurry videos, the existing video deblurring methods do not perform as well as the state-of-the-art image deblurring methods~\cite{Zamir2021MPRNet, Chen_2021_CVPR} on some benchmarks.
%
\newline
{\bf Real-World Blurry Datasets.} 
%
To better evaluate deblurring algorithms, some real-world blurry datasets~\cite{lai_blur, kohler, RealBlur, ESTRNN} have been proposed,
but all these works have their own drawbacks which limit their applications. 
In \cite{lai_blur}, Lai collects 100 real blurry images without paired sharp images, thus, their dataset is not suitable for quantitative evaluation.
\cite{kohler} proposes 48 pairs of real-world blurry images and corresponding sharp images through a high-precision hexapod robot, but the number of their dataset is too few for supervised training. 
%
Recently, ~\cite{RealBlur} proposes a real-world blurry dataset for single image deblurring, which contains 4554 pairs real-world blurry and sharp images of low-light static scenes.
Specifically, their device uses a beam splitter to separate the input light equally to these two cameras and an external synchronization trigger signal to activate the capture processes simultaneously.
With similar equipment and acquisition strategy, ~\cite{ESTRNN} proposes their Beam-Splitter Dataset (BSD) for video deblurring which consists of 100 paired real-world blurry and sharp videos. 
However, both datasets suffer from distinct misalignment due to their defective acquisition strategy.
%

\section{Proposed Algorithm}
Given consecutive blurry frames $I_{[t:t+n]}$, the goal of the proposed RNN-MBP method is to recover the corresponding sharp frames $O_{[t:t+n]}$ recurrently.
Figure \ref{fig:RNN} shows that the proposed RNN-MBP contains three modules:
\begin{compactitem}
    \item A feature extractor module to extract features from the target frame.
    \item A MBP module to gather the unaligned neighboring hidden states set and exploit the inter-frame information from different scales.
    \item A Target Frame Re-constructor module to reinforce the importance of intra-frame information and reconstruct sharp results.
\end{compactitem}

For a blurry frame $I_t$, our method first extracts target features $\Phi_t$ via a feature extractor that contains a convolutional layer and a Channel Attention Block~(CAB) from RCAN \cite{rcan} (Figure \ref{fig:RNN} (b)).
%
Then, a MBP module is developed to gather the extracted features $\Phi_t$ and unaligned neighboring multi-scale hidden states sets ($F_{t-1}$ and $B_{t+1}$) in both the forward direction and the backward direction, and then outputs the propagated multi-scale hidden state set ($F_{t}$ and $B_{t}$).
Finally, the Target Frame Re-constructor module progressively fuses the target features ($\Phi_t$) with the multi-scale hidden state sets ($F_{t}$ and $B_{t}$) and reconstruct the final deblurred results ($O_t$).

\subsection{MBP Module}
The proposed MBP is to exploit the inter-frame information by gathering target features and the unaligned neighbouring hidden states sets under a multi-scale framework.
To facilitate the gathering process, the proposed propagation module adopts two U-Net RNN cells and a multi-scale bidirectional propagation scheme (Figure~\ref{fig:RNN} (c)), where the unaligned neighbouring hidden states are integrated into different levels of the RNN cells. 
We adopt the multi-scale framework for two reasons:
(1) the blur degradations and the displacement among neighboring hidden states are relatively small in the features of lower resolution, which can help the cell to exploit the inter-frame information without explicit alignment, 
(2) the high-frequency information is still reserved in the features of larger resolution, which can help to reconstruct sharper results. 

%
The framework of one U-Net RNN cell can be divided into two parts: an encoder for gathering neighbouring hidden states and a decode for generating the propagated features.
Taking the forward propagation process as an example,
the encoder part in the forward U-Net RNN cell consists of three downsample operations and 6 CABs, which can progressively down-sample the target features $\Phi_t$ and gather them with the multi-scale hidden state set $F_{t-1}$ propagated from $t-1$ iteration.
Noted that the multi-scale hidden state set $F_{t-1}$ contains 6 hidden states and can be formulated as:
\begin{equation}\label{eqn:1}
F_{t-1} = (f^{e1}_{t-1}, f^{e2}_{t-1}, f^{e3}_{t-1}, f^{d1}_{t-1}, f^{d2}_{t-1}, f^{d3}_{t-1}),
\end{equation}
where $(f^{e1}_{t-1}, f^{e2}_{t-1}, f^{e3}_{t-1})$ and $(f^{d1}_{t-1}, f^{d2}_{t-1}, f^{d3}_{t-1})$ denote the hidden stats from different levels of the encoder and the decoder of the $t-1$ iteration RNN cell respectively.

Specifically, the gathering process in the different levels of the encoder can be formulated as:
\begin{equation}\label{eqn:2}
f^{e1}_{t} = \phi_\theta(\Phi_{t}) + Conv(f^{e1}_{t-1}) + Conv(f^{d1}_{t-1}),
\end{equation}
\begin{equation}\label{eqn:3}
f^{e2}_{t} = \phi_\theta((f^{e1}_{t})^{\downarrow 2}) + Conv(f^{e2}_{t-1}) + Conv(f^{d2}_{t-1}),
\end{equation}
\begin{equation}\label{eqn:4}
f^{e3}_{t} = \phi_\theta((f^{e2}_{t})^{\downarrow 2}) + Conv(f^{e3}_{t-1}) + Conv(f^{d3}_{t-1}),
\end{equation}
where the $\phi_\theta$ refers to two CABs, $Conv$ denotes a convolutional layer with filter size of $3 \times 3$, and the $\downarrow$ refers to down-sample operation.
Then, the decoder part progressively up-sample the features $f^{e3}_{t}$ and generate the propagated features with skip-connections from the encoder part ($f^{e2}_{t}, f^{e1}_{t}$),
which can be formulated as:
\begin{equation}\label{eqn:5}
f^{d3}_{t} = \phi_\theta(f^{e3}_{t})
\end{equation}
\begin{equation}\label{eqn:6}
f^{d2}_{t} = \phi_\theta((f^{d3}_{t})^{\uparrow 2}) + \phi_\theta(f^{e2}_{t})
\end{equation}
\begin{equation}\label{eqn:7}
f^{d1}_{t} = \phi_\theta((f^{d2}_{t})^{\uparrow 2}) + \phi_\theta(f^{e1}_{t})
\end{equation}
where $\uparrow$ denotes the up-sample operation.
Finally, all the six outputs in Eqs.~\eqref{eqn:2}-\eqref{eqn:7} will be grouped as the propagated multi-scale hidden state set $F_t$ and sent to the Target Frame Re-constructor and the next iteration.
Correspondingly, the backward propagation process also produce a multi-scale hidden state set $B_t$ which consists of $b^{e3}_{t}$, $b^{e2}_{t}$, $b^{e1}_{t}$, $b^{d1}_{t}$, $b^{d2}_{t}$, and $b^{d3}_{t}$.

\subsection{Target Frame Re-constructor}
\vspace{-1mm}
Although the inter-frame information is exploited in the MBP module, 
the intra-frame information and its importance may gradually vanish during the propagation process due to the introduction of the neighbouring hidden states.
Therefore, we propose a Target Frame Re-constructor module to re-introduce the target features ($\Phi_t$) and progressively fuse it with the propagated outputs from the forward and backward RNN cells ($F_t, B_t$).
We believe that the re-introduce of the target features can reinforce the importance of intra-frame information, which may further improve the performance of the re-constructor module.

The framework of the proposed Target Frame Re-constructor module is shown as Figure~\ref{fig:RNN} (c).
First, we up-sample all the hidden states to the same scale with target features and fused them level-by-level.
This process can be formulated as:
\begin{equation}\label{eqn:8}
f^1_{t} =  \psi _\theta(\Phi_t) + Conv(f^{e1}_{t}+b^{e1}_{t}) +  Conv(f^{d1}_{t}+b^{d1}_{t}),
\end{equation}
\begin{equation}\label{eqn:9}
f^2_{t} =  \psi_\theta(f^1_{t}) + (Conv(f^{e2}_{t}+b^{d2}_{t}))^{\uparrow 2} + (Conv(f^{d2}_{t}+b^{d2}_{t}))^{\uparrow 2},
\end{equation}
\begin{equation}\label{eqn:10}
f^3_{t} =  \psi_\theta(f^2_{t}) + (Conv(f^{e3}_{t}+b^{e3}_{t}))^{\uparrow 4} + (Conv(f^{d3}_{t}+b^{d3}_{t}))^{\uparrow 4},
\end{equation}
where $\psi_\theta$ denotes 8 CABs.
Finally, the deblurred result $O_t$ is restored by:
\begin{equation}\label{eqn:11}
O_t = Conv_{5*5}(f^3_{t}) + I_t,
\end{equation}
where $Conv_{5*5}$ denotes a convolutional layer with filter size of $5 \times 5$.
\section{Real-World Blurry Video Dataset}
\vspace{-0.5mm}
Due to the domain gap between synthetic blur and real blur~\cite{8953368}, models trained on synthetic blurry datasets perform not well on real-world blurry videos.
To collect a real-world deblurring dataset for both supervised training and evaluation, we propose a novel acquisition system to capture blurry videos and their corresponding ground truth.
The details of the acquisition system and acquisition process are described in the following sections.

\subsection{Digital Video Acquisition System}
\begin{figure}[!t] 
  \centering
  \includegraphics[width=0.9\linewidth]{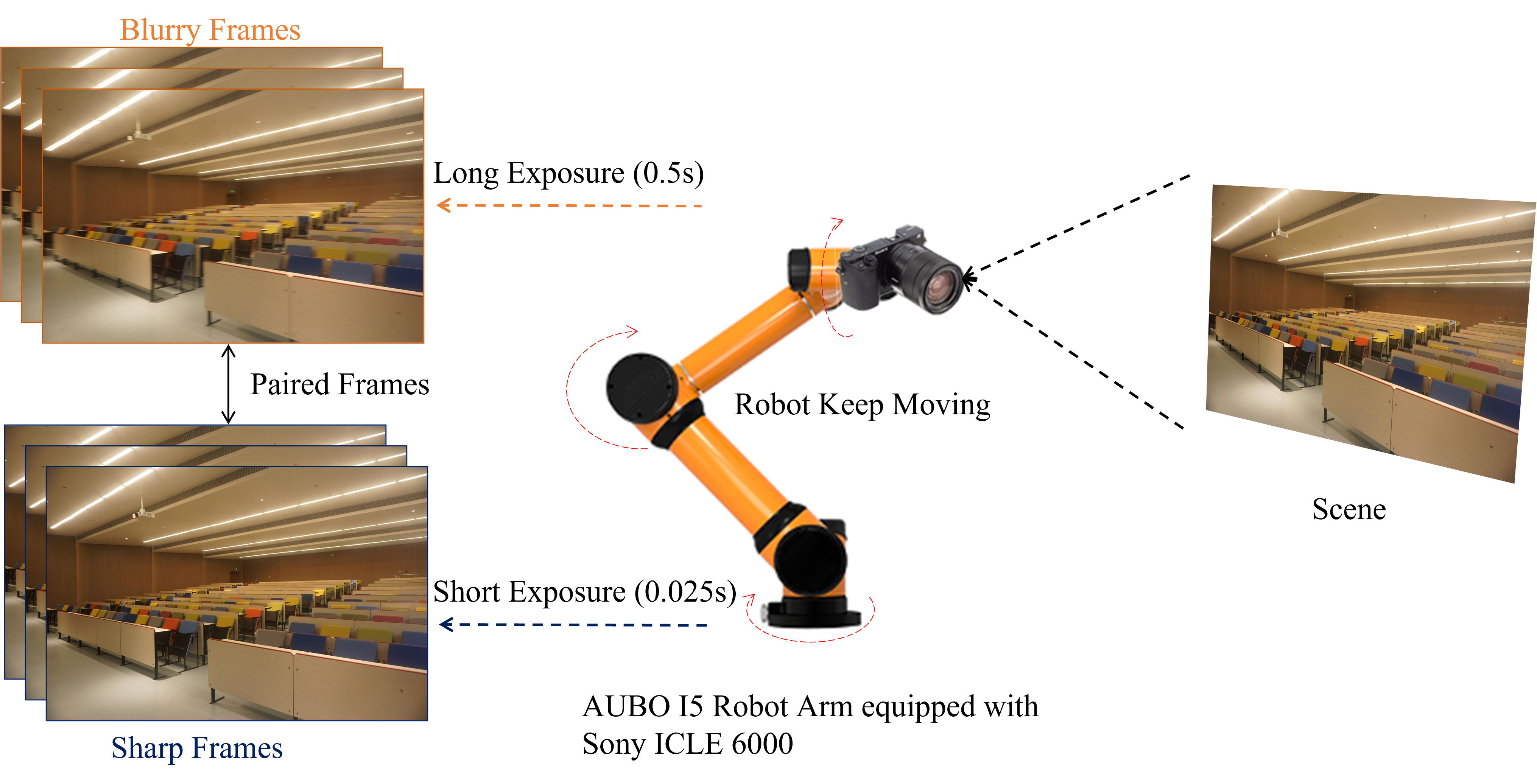}
  \vspace{-2mm}
  \caption{\textbf{Digital Video Acquisition System.}
}
\label{fig:VideoSystem}
\vspace{-4mm}
\end{figure}

In order to capture the paired real-world blurry video dataset ({RBVD}), we build our own Digital Video Acquisition System (DVAS).
Different from the acquisition process of \cite{ESTRNN} and \cite{RealBlur}, in which blurry and sharp images are captured simultaneously, the proposed DVAS can capture paired blurry and sharp videos respectively.
As shown in Figure~\ref{fig:VideoSystem}, our system consists of four components, 
a high accurate robot arm (AUBO I5), 
a controller case of the robot arm, a Sony ICLE 6000 camera, 
and a laptop to tele-control the camera. 
Since the AUBO robot arm can repeatedly reach the same location with an accuracy of 2mm, we can collect perfectly aligned blurry and sharp video frames at a series of locations. 
%
%
%

\subsection{Dataset Acquisition}
{\bf Scene layout and system settings.} 
To guarantee the generality of the proposed dataset, all scenes in our dataset have been carefully selected.
Scenes with rich and colorful textures are preferred as the primary goal of deblurring is to recover high-frequency information from blurry inputs.
Specifically, our dataset contains 40 indoor scenes, 14 outdoor scene, and 9 scenes with high-frequency charts. 
%


%
Before running the acquisition process, we first fix the remote-controlled camera at the end of the robot arm by a customized mounting rack and set the velocity of the robot arm to less than 0.25m/s to avoid the undesired shake of camera.
Then, the camera's exposure time is set to 0.5s and 0.025s to collect blurry videos and corresponding sharp videos, respectively. 
Since capturing blurry videos requires a long exposure time, we use a neutral density filter (ND5) to eliminate overexposure problems.
To enrich the diversity of our dataset, we also design 16 pre-defined trajectories for the following acquisition process.
More details about the optical parameters of the camera and robot arm settings can be found in the supplementary material.
\newline
{\bf Acquisition process.}  
%
\begin{figure}[!t] 
  \centering
  \includegraphics[width=0.9\linewidth]{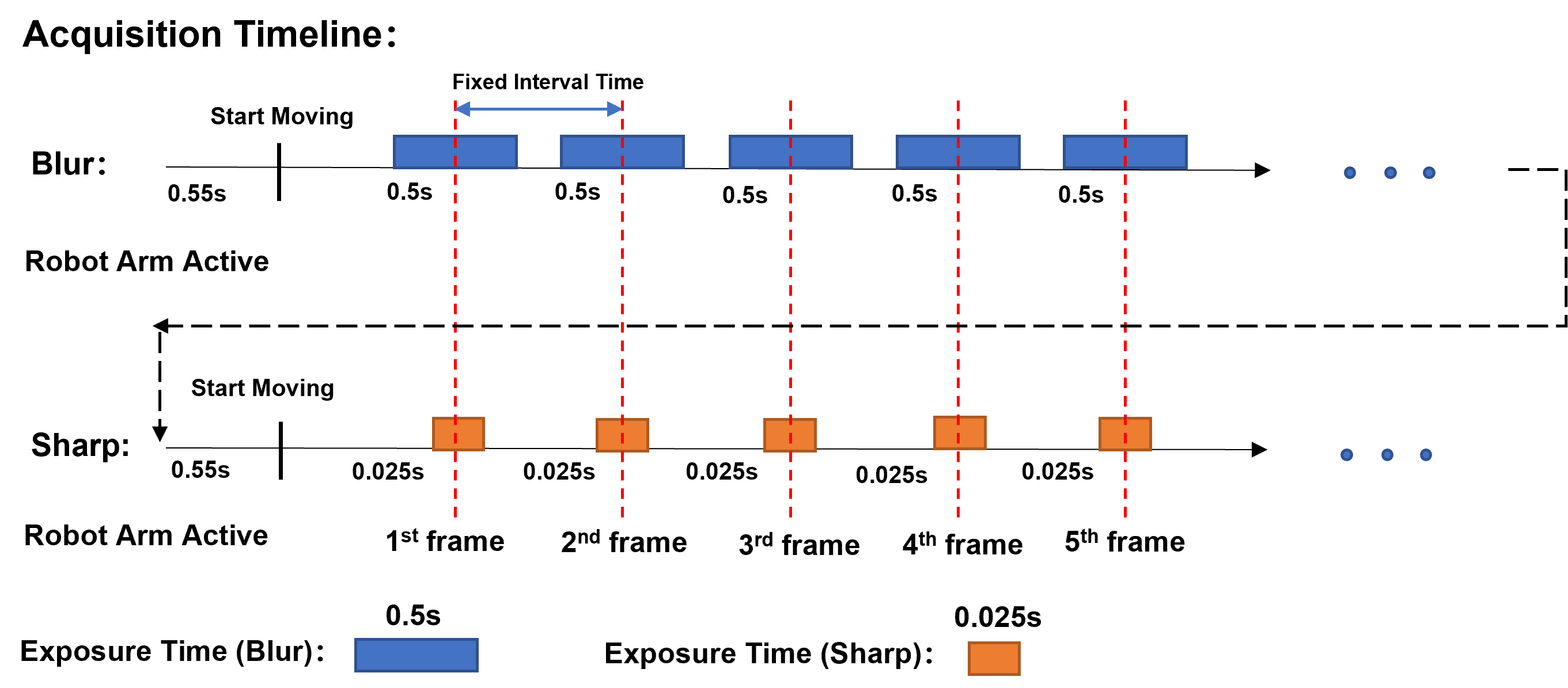}
  \vspace{-2mm}
  \caption{\textbf{Timeline of our acquisition process.} To capture the pairs of blurry and corresponding sharp videos, the camera will move twice along a pre-defined trajectory.}
\label{fig:Video_Timeline}
\vspace{-6mm}
\end{figure}
%
After finishing all the preparatory works, we start a multi-threading program to pick a pre-defined trajectory and control the robot arm moving to a series of acquisition locations.
The timeline of the acquisition process can be summarized as Figure \ref{fig:Video_Timeline}.
The program takes 0.55 second to initialize the robot arm and the acquisition program starts at 3 seconds after robot arm moves.
During the movement, the multi-threading program controls the camera taking blurry frames at the desired timing according to the exposure time.
Depending on which trajectories we used in the acquisition process, the number of acquisition points ranges from 30-60, which means one captured video may contain 30-60 consecutive frames.
The blurry video acquisition process usually takes approximately 5 minutes, and then the robot arm is re-initialized to the starting point for capturing the sharp video.
It is noted that the centers of the exposure time for two acquisition processes are coincided to guarantee the spatial alignment of the corresponding blurry and sharp frame. 
Finally, all the collected video pairs are post-processed (correction, cropping and downsampling) to get suitable size (1440 $\times$ 960) and perfectly aligned video for training and evaluation.

\begin{table*}[!t] 
\centering
\begin{adjustbox}{width=0.95\linewidth}
\begin{tabular}{c|c c c c c c c c c c c c}
    \hline
    {\bf Methods}  & EDVR     & DVD-HC      & DeblurGAN-v2    & ESTRNN     & DMPHN    & CDVD-TSP   & SPN           & MPRNet   & HINet    & RNN-MBP     \\
    \hline
    Params         & 23.6M    & 15.3M       & 60.9M           & 5.6M       & 21.7M    & 16.2M      & 23.0M         & 20.1M    & 88.7M    & 16.4M    \\
    Time           & 1.11s    & -           & 1.27s           & 0.23s      & 2.49s    & 7.80s      & -             & 2.50s    & 2.18s    & 1.12s     \\
    PSNR           & 26.83    & 27.31       & 29.55           & 31.07      & 31.2     & 31.67      & 31.85         & 32.66    & 32.71    & {\bf 33.32}    \\
    SSIM           & 0.8426   & 0.8255      & 0.934           & 0.9023     & 0.940    & 0.9279     & 0.948         & 0.9590   & 0.9590   & {\bf 0.9627}   \\
    \hline
\end{tabular}
\end{adjustbox}
\vspace{-2mm}
\caption{{\bf Quantitative evaluations on the GORPO dataset.}}
\label{goproTable}
\vspace{-3mm}
\end{table*}
\begin{figure*}[!t] 
  \centering
  \includegraphics[width=0.9\linewidth]{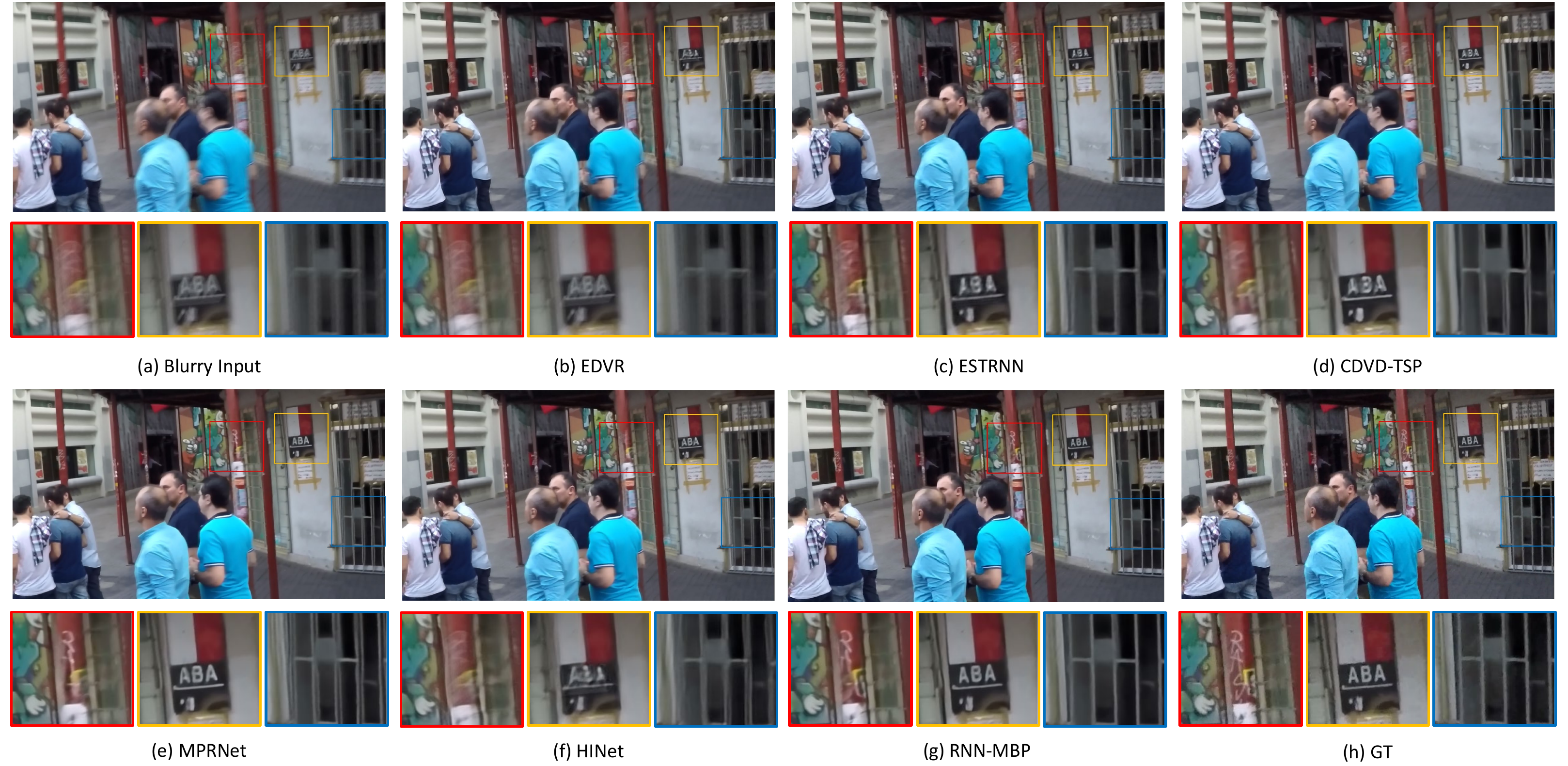}
  \vspace{-4mm}
  \caption{\textbf{Visual results on the GOPRO dataset.} The results in (b)-(f) contain blur residual and ghost artifacts, while the deblurred image in (g) by our method is much clearer. Best viewed on a high-resolution display.}
\label{fig:goproResult}
\end{figure*}

%
\begin{table*}[!t] 
\centering
\vspace{-2mm}
\begin{adjustbox}{width=0.90\linewidth}
\begin{tabular}{c|c c c c c c c c c c c}
    \hline
    {\bf Methods}  & GVD-DS          & DL-HMB        & SRNN         & DVD-HC      & DTBN         & EDVR      & STFAN    & CDVD-TSP    & RNN-MBP           \\
    \hline  
    PSNR           & 26.94           & 28.27         & 29.98        & 30.01       & 29.95        & 28.51     & 31.15    & 32.13       & {\bf 32.49}    \\
    SSIM           & 0.8158          & 0.8463        & 0.8842       & 0.8877      & 0.8692       & 0.8637    & 0.9049   & 0.9268      & {\bf 0.9568}   \\
    \hline
\end{tabular}
\end{adjustbox}
\vspace{-2mm}
\caption{{\bf Quantitative evaluations on the dataset by~\cite{Su_2017_CVPR}.} Instead of randomly selected 30 frames from each dataset of ~\cite{Su_2017_CVPR}, all frames of the test set are used for evaluation.}
\label{DVDTable}
\vspace{-2mm}
\end{table*}

\section{Experiments}
\subsection{Training Dataset and Details}
We use two synthetic datasets~(GOPRO~\cite{deepdeblur} and dataset by~\cite{Su_2017_CVPR}) and the proposed RBVD dataset in our experiments.
The proposed RBVD dataset contains 63 video sequences (2164 samples in total): 7 video sequences are divided into the test set (including 2 outdoor scenes, 4 indoor scenes, and 1 scene with high-frequency charts), and the remaining 56 sequences are used as the training set (including 12 outdoor scenes, 36 indoor scenes and 8 scenes with high-frequency charts). 

For each dataset, we train the proposed model using its training set and evaluate it on the corresponding test set.
In the training process, we adopt L1 Charbonnier loss and Adam optimizer~\cite{2014Adam} with default parameters to train our model.
The learning rate is initialized as $2\times10^{-4}$ and is updated with the Cosine Scheduler strategy~\cite{2016SGDR}. 
We set the length of the input sequences to 8 as a trade-off between the performance and required GPU memory. 
The batch size is 4 and the patch size of input sequences is 256 $\times$ 256.
%
The proposed model is implemented by the PyTorch framework and trained on 4 NVIDIA Tesla V100 GPUs.
The code and the proposed RBVD dataset will be made publicly available.
%
%

\begin{figure*}[!t] 
  \centering
  \includegraphics[width=0.95\linewidth]{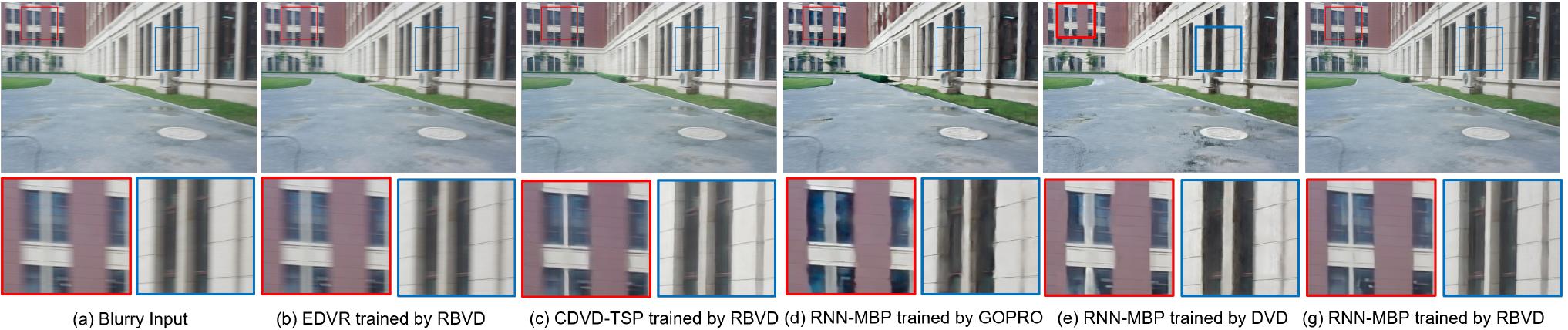}
  \vspace{-3mm}
  \caption{\textbf{Deblurred results on test set of the RBVD dataset.} Best viewed on a high-resolution display.}
\label{fig:RBVDresults}
\vspace{-5mm}
\end{figure*}

\subsection{Performance Evaluation}
{\bf Evaluations on the synthetic datasets.} 
We evaluate different video deblurring algorithms on two synthetic datasets (GOPRO~\cite{deepdeblur} and dataset by~\cite{Su_2017_CVPR}) and use the PSNR and the SSIM as the evaluation metrics.
For fair comparisons, all the deep learning-based methods are trained on the same settings for evaluation with the same machine (Nvidia RTX 8000).

For the GOPRO dataset, the evaluated methods include both state-of-the-art video deblurring methods (EDVR~\cite{2019EDVR}, DVD-HC~\cite{Su_2017_CVPR}, ESTRNN~\cite{ESTRNN}, CDVD-TSP~\cite{Pan_2020_CVPR}) and image deblurring methods (DMPHN~\cite{8953205}, SPN~\cite{2020Spatially}, MPRNet~\cite{Zamir2021MPRNet}, HINet~\cite{Chen_2021_CVPR}).
%
Table~\ref{goproTable} shows the quantitative results on the GOPRO~\cite{deepdeblur}.
We note that the state-of-the-art video deblurring methods~\cite{2019EDVR, Pan_2020_CVPR} do not perform well compared with the image deblurring methods~\cite{Zamir2021MPRNet, Chen_2021_CVPR} due to the inaccurate estimations of the alignment. 
In contrast, our method generates the best results and its PSNR value is at least 0.61dB higher than the evaluated methods. 
This result demonstrates that the proposed method can effectively exploit intra- and inter-frame information by directly gathering the unaligned neighboring hidden states.
In addition, the proposed network involves fewer model parameters and fewer running time, which further demonstrate its effectiveness.
Figure \ref{fig:goproResult} shows an example of the visual results. We note that the evaluated methods produce ghost artifacts on the iron fence of the door (the blue box in Figure \ref{fig:goproResult}).
In contrast, the proposed RNN-MBP generates much clearer images with sharper characters, clear graffiti on the pillar, and correct iron fence (Figure \ref{fig:goproResult}(g)).

We further compare our algorithm with several state-of-the-art video deblurring algorithms including GVD-DS~\cite{7299181}, DL-HMB~\cite{8099888}, SRNN~\cite{8578951},  DVD-HC~\cite{Su_2017_CVPR}, DTBN~\cite{Kim_2017_ICCV}, EDVR~\cite{2019EDVR}, STFAN~\cite{9010007}, and CDVD-TSP~\cite{Pan_2020_CVPR}.    
Table~\ref{DVDTable} shows the quantitative results on the dataset by~\cite{Su_2017_CVPR}.
The proposed RNN-MBP achieves the most competitive results against the state-of-the-art video deblurring methods (by a margin of 0.36db) due to its effectiveness on exploiting the inter- and intra-frame information. 
More qualitative comparisons are included in the supplementary material. 
%
\begin{figure}[!t] 
  \centering
  \includegraphics[width=0.85\linewidth]{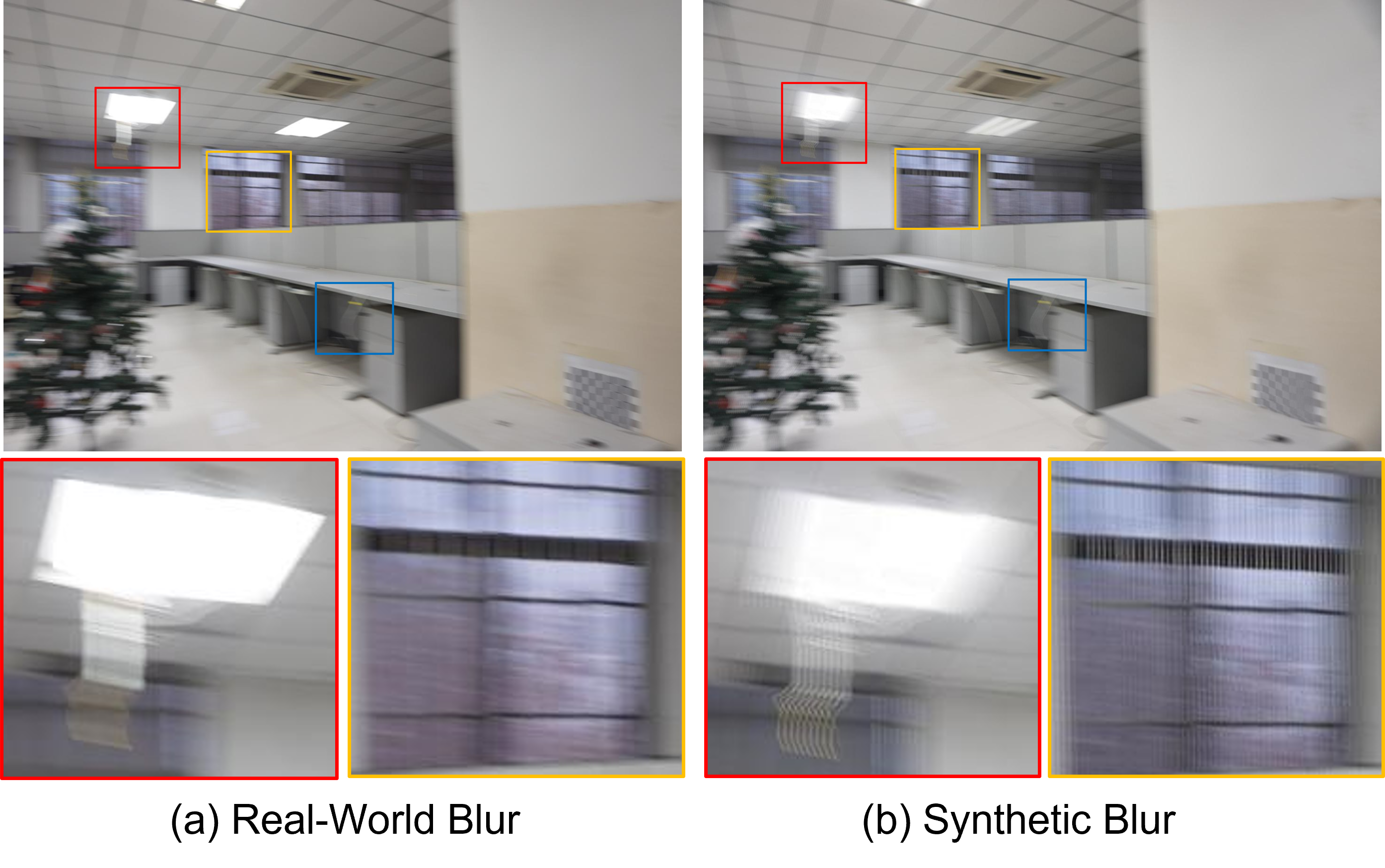}
    \vspace{-3mm}
  \caption{\textbf{The difference between real blur (a) and synthetic blur (b).} Best viewed on a high-resolution display.}
\label{fig:synVSreal}
\vspace{-2mm}
\end{figure}
%

{ \bf Evaluations on the RBVD.} Due to the domain gap between synthetic blur and real-world blur~\cite{8953368}, the methods trained on the synthetic datasets do not perform well on the real-world ones.
%
To better illustrate the difference between the real-world blur and synthetic blur, we generate a synthetic blurry video by averaging a series of the clear scenes from a video in the proposed RBVD according to the protocol of~\cite{deepdeblur} and compare the generated blurry videos with our real captured blurry videos.
Figure~\ref{fig:synVSreal} shows that the synthetic blurry frame contains significant discontinuous artifacts while the blurry frame collected by DVAS is more realistic. 
Therefore, the model trained on such synthetic blurry datasets may not perform well on the real-world blurry videos. 
%

To demonstrate whether the proposed RBVD dataset facilitates the performance of existing deblurring methods, we train and evaluate the proposed RNN-MBP and two representative video deblurring methods \cite{2019EDVR, Pan_2020_CVPR} on the proposed RBVD dataset. 
In addition, we also evaluate our method trained on the GOPRO and dataset by~\cite{Su_2017_CVPR} using the test set of the RBVD.
%
Table~\ref{RBVDTable} shows that the proposed RNN-MBP performs better than EDVR~\cite{2019EDVR} and CDVD-TSP~\cite{Pan_2020_CVPR} with a marked improvement of 0.589dB and 1.115dB in terms of PSNR.
In addition, the huge performance gaps among the RNN-MBP models trained on the RBVD dataset and synthetic datasets demonstrate the training set of the proposed RBVD is more suitable for training a real-world oriented deblurring models.
Figure~\ref{fig:RBVDresults} shows an example of the visual results on the RBVD dataset.
As shown in Figure~\ref{fig:RBVDresults} (e), only the RNN-MBP trained on the RBVD can generate clear results with sufficient details, while other deblurring methods and the RNN-MBP trained on the synthetic datasets suffer from irregular edges and blurry results.
\begin{table}[!t] 
\centering
\begin{adjustbox}{width=\linewidth}
\begin{tabular}{c c c c c c}
    \hline
    {\bf Methods}   & EDVR     & CDVD-TSP    & RNN-MBP(GOPRO)    & RNN-MBP (DVD)   & RNN-MBP   \\
    \hline
    PSNR            & 25.966   & 25.44       & 22.826            & 23.744         & 26.555   \\
    SSIM            & 0.8285   & 0.8167      & 0.8662            & 0.8741         & 0.9016    \\ 
    \hline
\end{tabular}
\end{adjustbox}
\vspace{-2mm}
\caption{{\bf Quantitative and qualitative evaluations on the RBVD dataset.} Models are trained on the RBVD train set unless mentioned specially.}
\label{RBVDTable}
\vspace{-5mm}
\end{table}

\vspace{-4mm}
\subsection{Ablation Studies}
We demonstrate the effectiveness of each key component in the proposed RNN-MBP.
All the methods mentioned below are trained on the GOPRO~\cite{deepdeblur} dataset with the same settings for fairness. 
Since the BasicVSR~\cite{chan2021basicvsr} adopts RNN architecture and achieves promising results in the video super-resolution task, we remove the upsampling module and flow-based alignment module to form a basic RNN for video deblurring, which denotes as Baseline.
%
%
Then, to better exploit inter-frame information from the neighboring hidden states, we replace the RNN cell in the Baseline with the proposed MBP module.
%
We also introduce explicit alignments modules to the Baseline, denoted as Baseline+SpyNet and Baseline+deformable respectively, to evaluate the influence of explicit alignments in the video deblurring task.
%
%
%
%
%
Finally, we deploy the proposed Target Frame Re-constructor module in the Baseline+MBP model to reinforce the intra-frame from the target features, which form our final model RNN-MBP.

Table~\ref{ablation} shows the effectiveness of the proposed MBP and Target Frame Re-constructor module.
Compared with the state-of-the-art image deblurring algorithms, the Baseline+SpyNet and Baseline+deformable cannot achieve promising results due to the inaccurate alignments.
The introduction of the MBP module achieves 1.398 dB performance improvement over the Baseline+SpyNet model, which verifies that directly gathering the unaligned neighboring hidden states with a multi-scale propagation scheme is effective in exploiting inter-frame information.
Although the Baseline+deformable contains an extra alignment module, it shows much worse performance (5.853dB) than the Baseline+MBP, which demonstrates that the unsatisfying warped features will deteriorate the performance of the video deblurring methods. 
%
Applying the proposed Target Frame Re-constructor module can further exploit the intra-information by re-introducing the target features, which outperforms the Baseline+MBP model by a margin of 1.232 dB.
Overall, the success of the RNN-MBP demonstrates that a well-designed recurrent neural network with the multi-scale propagation scheme is sufficient for exploiting the inter- and intra-frame information and obtaining favorable results.
%

\begin{table}[!h] 
\vspace{-3mm}
\centering
\begin{adjustbox}{width=\linewidth}
\begin{tabular}{r|c c c | c c c}
    \hline
    {\bf Model}           & Alignment                      & MBP          & TFR        & Parameters      & PSNR       & SSIM     \\
    \hline
    Baseline              &                                &              &            & 10.24M          & 31.443     & 0.9442   \\
    Baseline+MBP          &                                & \checkmark   &            & 10.35M          & 32.087     & 0.9524   \\
    Baseline+SpyNet       & \checkmark(Spy-Net)            &              &            & 11.68M          & 30.689     & 0.9387   \\
    Baseline+deformable   & \checkmark(deformable layers)  &              &            & 10.60M          & 26.234     & 0.8322   \\         
    RNN-MBP               &                                & \checkmark   & \checkmark & 16.37M          & 33.319     & 0.9627   \\
    \hline
\end{tabular}
\end{adjustbox}
\vspace{-2mm}
\caption{{\bf The ablation study of the proposed algorithm.} All the methods are trained on the GOPRO dataset using the same training settings.}
\label{ablation}
\vspace{-5mm}
\end{table}

\vspace{-1mm}
\section{Conclusion}
In this paper, we proposed an effective Recurrent Neural Network with Multi-scale Bi-directional Propagation (RNN-MBP) for video deblurring.
By adopting the U-Net RNN cells and multi-scale bidirectional propagation scheme, 
the proposed RNN-MBP can effectively exploit the inter-frame information from the unaligned neighboring hidden states without explicit alignment.
In addition, we also developed a Target Frame Re-constructor module to reinforce the importance of the intra-frame information by re-introducing the target features into the reconstruction process.
The ablation studies verify that the proposed modules are effective for video deblurring.
To better evaluate the performance of video deblurring methods on real-world scenes, we also collect a Real-World Blurry Video Dataset (RBVD) by a well-designed Digital Video Acquisition System (DVAS).
Extensive evaluations on the synthetic and real-world datasets show that the proposed RNN-MBP performs favorably against the state-of-the-art deblurring methods.

%

\section{Acknowledgments}
C. Zhu, B. Liang, Y. Huang and F. Wang are supported in part by National Major Science and Technology Projects of China (No.2019ZX01008101), the Fundamental Research Funds for the Central Universities (FRFCU), the Natural Science Foundation of Shaanxi Province (CN) (2021JQ-05). 
%
%
J. Pan is supported in part by NSFC (Nos. 61872421,61922043), the FRFCU (No. 30920041109).

\bibliography{aaai22}

\end{document}